\documentclass[prb,preprint,amsmath,amssymb]{revtex4-1}
\usepackage[utf8x]{inputenc}
\usepackage{graphicx}
\usepackage{float}
\usepackage{dcolumn}% Align table columns on decimal point
\usepackage{bm}% bold math
\usepackage{amssymb}
\usepackage{epsfig}
\usepackage{psfrag}
\usepackage{amsfonts}
\begin{document}
%\linenumbers
 %\begin{frontmatter}

\title{Pressure-induced phase transition in Bi$_2$Se$_3$ at 3 GPa: electronic topological transition or not? }

\author{Achintya Bera$^1$, Koushik Pal$^{2}$, D V S Muthu$^1$, U V Waghmare$^2$ and A K Sood$^1$,\footnote{electronic mail:asood@physics.iisc.ernet.in}}

\address{$^1$Department of Physics, Indian Institute of Science, Bangalore 560 012, India}
\address{$^2$ Chemistry and Physics of Materials Unit and Theoretical Sciences Unit, Jawaharlal Nehru Centre for Advanced Scientific Research, Bangalore-560064, India}

\date{\today}

\begin{abstract}
In recent years, a low pressure transition around P $\sim$ 3 GPa exhibited by the A$_2$B$_3$-type 3D topological insulators is attributed to an electronic topological transition (ETT) for which there is no direct evidence either from theory or experiments. We address this phase transition and other transitions at higher pressure in bismuth selenide (Bi$_2$Se$_3$) using Raman spectroscopy at pressure upto 26.2 GPa. We see clear Raman signatures of an isostructural phase transition at P $\sim$ 2.4 GPa followed by structural transitions at $\sim$ 10 GPa and 16 GPa. First-principles calculations reveal anomalously sharp changes in the structural parameters like the internal angle of the rhombohedral unit cell  with a minimum in the c/a ratio near P $\sim$ 3 GPa. While our calculations reveal the associated anomalies in vibrational frequencies and electronic bandgap, the calculated $\mathbb{Z}_2$ invariant and Dirac conical surface electronic structure remain unchanged, showing that there is no change in the electronic topology at the lowest pressure transition.
\end{abstract}

%\noindent{\it Keywords\/}: Raman, high pressure, topological insulator, electronic topological transition, Lifshitz transition

%\end{frontmatter}

%\linenumbers
\maketitle
%\begin{multicols}{2}

%\narrowtext
\section{Introduction}
\label{sec:introduction}

Recently, a lot of research has focused on the novel physics of topological insulators (TIs) \cite{yando, mfru, moore, hasan, qi, fu, mele, hug} which exhibit metallic surface states protected by the time reversal symmetry (TRS) or by different point group symmetries of the underlying crystal such as crystalline TI \cite{dzi}. Among the TRS protected topological insulators, Bi$_2$Se$_3$ shows a lot of promise for unique applications such as, in quantum magnetoresistance \cite{yang}, generation of highly polarized spin currents \cite{wang}, low power electronics \cite{zutic}, quantum computation, \cite{nayak, more} etc. Because of its layered structure, different methods such as mechanical exfoliation by scotch tape, by an atomic force microscope tip \cite{hong}, chemical vapour transport and a polyol method \cite{zha}, are being used to prepare a few quintuple layers of Bi$_2$Se$_3$ for transport experiments.

The presence of single Dirac cone in the surface electronic structure of Bi$_2$Se$_3$ is the hallmark of TIs \cite{yzhang2}. However, these surface states are not easy to probe experimentally as they can be masked by the contribution of the bulk carriers to the measured conductivity. Transport measurements have been performed on Bi$_2$Se$_3$ to explore the surface states \cite{hstein1, kim1} by observing V-shaped conductance, ambipolar behaviour and the enhancement in thermopower \cite{kim2} due to Dirac electrons. Spin-orbit coupling (SOC) plays a crucial role in realization of TIs, like Bi$_2$Se$_3$, Bi$_2$Te$_3$ and Sb$_2$Te$_3$. By calculating the parities of the occupied bands of Bi$_2$Se$_3$, Zhang et al. showed \cite{hzhang} that the SOC leads to the reversal of valence and conduction bands of opposite parity, and hence changes the $\mathbb{Z}_2$ topological index from 0 to 1, characterizing the electronic topological transition (ETT).  Mechanical strain also has an impact on the topological insulating phase by tuning the relative strength of SOC vis-a-vis the crystal field splitting \cite{pal2014strain}. It was shown that application of hydrostatic pressure ($\sim$ 3 GPa) turns a band insulator Sb$_2$Se$_3$ into a topological insulator \cite{PRL1}, whereas a topological insulator Bi$_2$Se$_3$ can be converted into a band insulator by applying tensile strain \cite{steve, liu1}.

An iso-structural compound, Bi$_2$Te$_3$ undergoes a superconductivity transition \cite{einaga, jl, czhang1} with pressure around 4 GPa (at low temperature), providing a platform to probe Majorana Fermionic state \cite{andy} existing at the superconductor/TI interface \cite{ewang}. The enhancement in thermoelectric power \cite{svovs1} of Bi$_2$Te$_3$, increase in resistivity  and monotonous decrease in mobility for Bi$_2$Se$_3$ have been observed in the $\sim$ 3-5 GPa pressure range \cite{hamlin}. A change in the bulk modulus parallel to the layers (and a minimum in the c/a ratio) in the  3-5 GPa pressure range has been observed for all three stoichiometric TIs  Bi$_2$Se$_3$ \cite{bs}, Bi$_2$Te$_3$ \cite{bt} and Sb$_2$Te$_3$ \cite{st}.  This low pressure transition has been termed as electronic topological transition \cite{poli} (ETT) or Lifshitz transition \cite{lif}, without explicitly demonstrating any changes in the electronic topology.

Here, it is worthy to define ETT which can have two implications: (a) The van Hove singularity associated with the band extrema passes through the Fermi level, and thereby the distribution of carriers and Fermi surface topology changes. This is also known as Lifshitz \cite{lif} transition assigned for all the aforementioned TIs \cite{bs,bt,st,gopal,PSSB_Review} in the range of 3-5 GPa; (b) another type of ETT is characterized by the $\mathbb{Z}_2$ topological index, when the reversal of valence band and the conduction band with opposite parities occurs and as a result, odd number of surface Dirac cones appear \cite{hzhang,PRL1,PRL_support}. We revisit this low pressure transition in the present work using high pressure Raman experiments along with first-principles calculations on Bi$_2$Se$_3$. We do not find any change in electronic topology of the both types mentioned above as a function of pressure (P $\leq$ 8 GPa) by examining the density of states at Fermi level and the smallest electronic band gap as well as the $\mathbb{Z}_2$ index and the surface Dirac conical electronic structure. The pressure derivatives of Raman modes show a clear change at 2.4 GPa, without appearance of any new mode. The angle of the rhombohedral unit cell increases sharply upto $\sim$ 3 GPa followed by a slow increase. We term the 3 GPa transition as iso-structural transition (IST), instead of ETT.

Raman spectroscopic measurements are quite effective in probing phase transitions in topological insulators with pressure either by change in pressure coefficients or in full width at half maximum (FWHM) of phonons or appearance of new modes, as seen in Sb$_2$Se$_3$ \cite{PRL1} and in Bi$_2$Te$_3$ \cite{gopal}. On account of differences (discussed later) with results reported by Vilaplana et al. \cite{bs}, here we report high pressure Raman study on Bi$_2$Se$_3$ single crystal upto 26 GPa to benchmark our methods through observation of the other structural-phase transitions i.e. $\alpha\rightarrow\beta$ at $\sim$ 10 GPa, $\beta\rightarrow\gamma$ at $\sim$ 16 GPa and $\gamma\rightarrow\delta$ at $\sim$ 25 GPa respectively, and then focus on the IST at P $\sim$ 3 GPa.

\section{Experimental Details}
\label{sec:experimental Details}

Thin platelets ($\sim$ 30-40 $\mu$m thick ) cleaved from Bi$_2$Se$_3$ single crystals were placed together with a ruby chip into a stainless steel gasket inserted between the diamonds of a Mao-Bell-type diamond anvil cell (DAC). Methanol-ethanol (4:1) mixture was used as the pressure transmitting medium; the pressure was determined via the ruby fluorescence shift. Unpolarized Raman spectra were recorded in backscattering geometry using 488 nm excitation from an Ar$^+$ ion laser (Coherent Innova 300). The spectra were collected by a DILOR XY Raman spectrometer coupled to a liquid nitrogen cooled charged coupled device, (CCD 3000 Jobin Yvon-SPEX). After each Raman measurement, calibration spectra of a Ne lamp were recorded to correct for small drifts, if any, in the energy calibration of the spectrometer. Laser power($<$ 5 mW) was held low enough to avoid heating of the sample. The peak positions were determined by fitting Lorentzian line shapes with an appropriate background.

\section{Computational Details}

We employ the {\sc Quantum ESPRESSO} (QE) code \cite{GIANOZZI} for first-principles calculations based on density functional theory along with both fully relativistic and scalar-relativistic pseudopotentials.  While the spin-orbit coupling (SOC) has profound effect in modifying electronic properties of a material containing heavy elements, it has only weak effects on  vibrational frequencies due to different energy scales associated with spin-orbit coupling and vibrational energy of the ions. As a consequence, the magnitude of the phonon frequencies does not change significantly and the character of the  vibrational modes also remains unchanged with omission of the SOC. Effect of spin-orbit coupling is included in our first-principles calculations by constructing relativistic pseudopotentials from the solution of the Dirac equation instead of non-relativistic Schr\"{o}dinger equation. We have used relativistic pseudopotentials in determination of electronic structure of Bi$_2$Se$_3$ and scalar-relativistic (average of the former) pseudopotentials for calculating its vibrational properties. By scalar relativistic pseudopotentials,  we mean that the pseudopotentials are constructed taking into account the mass-velocity correction term and the Darwin term in the  higher order (in order of v$^{2}$/c$^{2}$, where v and c are velocity of electron and speed of light, respectively) expansion of the Dirac equation while excluding the spin-orbit coupling term. We adopt the self-consistent  density functional perturbation theory (DFPT) \cite{baroni} available within the QE distribution in calculation of vibrational frequencies.  We approximate the exchange-correlation energy with a generalized-gradient approximation (GGA)\cite{HUA} and use the functional parameterized by Perdew, Burke and Ernzerhof (PBE) \cite{PERDEW}. The kinetic energy cut-offs on the plane wave basis for the wave function and charge density are kept at 60 Ry and 240 Ry, respectively.  Integrations over the Brillouin zone are performed with a dense mesh of $9\times9\times9$ k-points \cite{PACK} while calculating the bulk electronic structure of the $\alpha$ phase. Enthalpies for the three different phases ($\alpha$, $\beta$ and $\gamma$) of Bi$_2$Se$_3$ are calculated at appropriately chosen k-meshes. Occupation numbers are treated according to the Fermi-Dirac distribution function with a broadening width of 0.003 Ry. We allowed the relaxation of the cell parameters as well as of the atomic positions in the  bulk unit cell at each target  pressure (hydrostatic) until the magnitude of forces on atoms at each pressure are less than 1 mRy/bohr.

Bi$_2$Se$_3$ is a strong 3D topological insulator with a layered structure having R$\bar{3}$m space group \cite{wise} and $D_{3d}$ point group symmetries. The rhombohedral unit cell of the bulk crystal contains five atomic planes with  Se1-Bi-Se2-Bi-Se1 stacking in a quintuple layer (QL) along $\langle111\rangle$ direction. Two inequivalent Se atoms (Se1 and Se2) occupy the outermost and the central planes of a QL respectively, and Se2 site in the central plane is an inversion center of the crystal structure.

The surface electronic structure of Bi$_2$Se$_3$ is calculated with a 10$\times$10$\times$1 mesh of k-points in the Brillouin zone of a hexagonal supercell containing a slab of QLs stacked along the z-direction. While constructing the slab for the surface calculations, we use the relaxed coordinates of the bulk and use 6 QLs above which a vacuum of 15 $\AA{}$ is added to prevent the electrostatic interaction between the periodic images of the slab under consideration.

\section{Results and Discussion}
\label{sec:Results and Discussion}

\begin{figure}[p!]
\centering
\includegraphics[trim=0 0 40 20, scale=0.6]{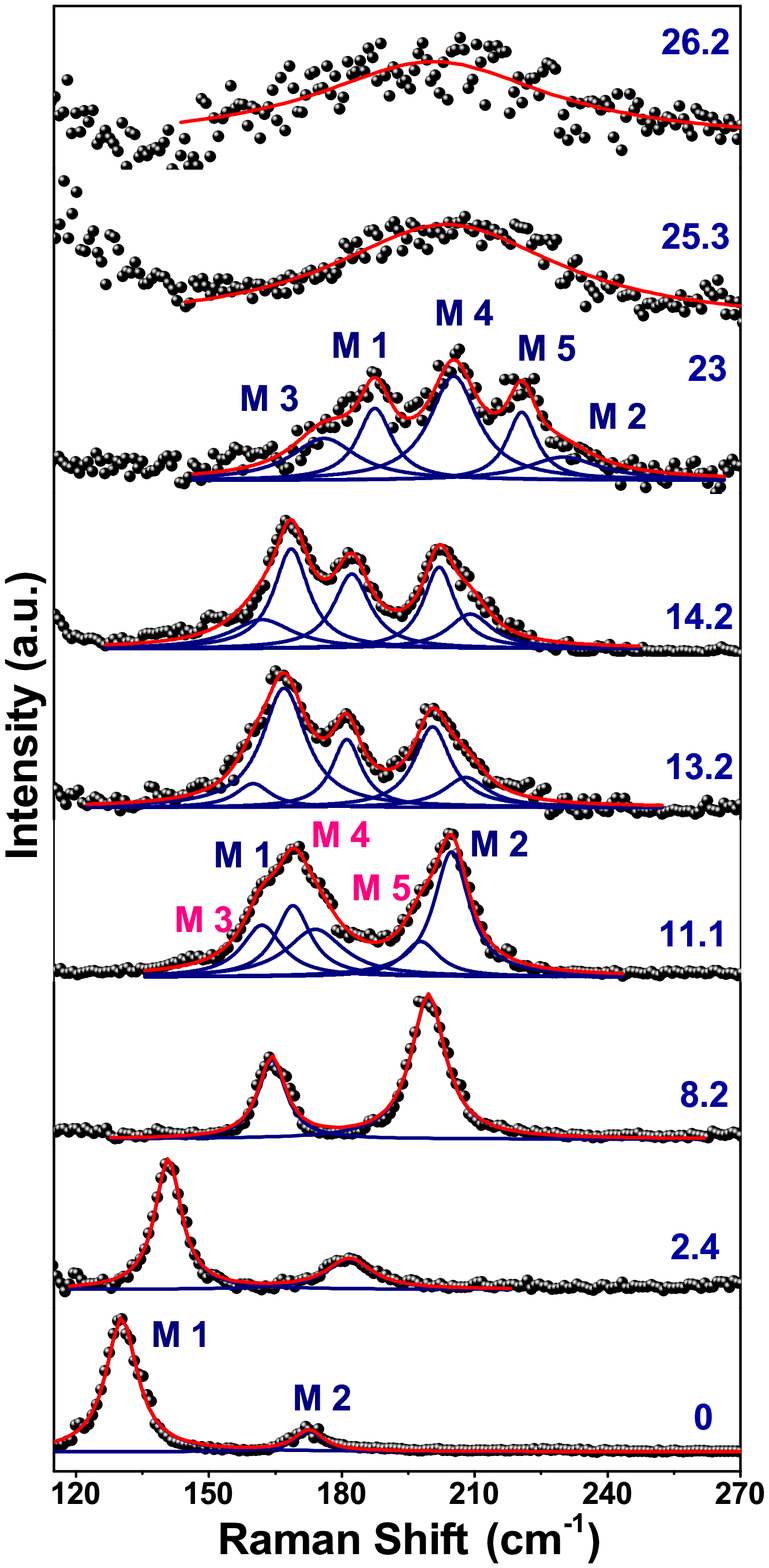}
\caption{(color online)-- Pressure evolution of Raman spectra. The solid lines are Lorentzian fits to the experimental data points (solid circles). New modes are marked as M3, M4 and M5.}
\label{Fig_1}
\end{figure}

The crystal structure of bulk Bi$_2$Se$_3$ is rhombohedral (R$\bar{3}$m) \cite{wise} with experimental lattice parameters a=4.143 $\AA$ and c=28.636 $\AA$ \cite{nakajima}. Experimental value of the band gap is $\sim$ 0.35 eV \cite{black} whereas estimate of the band gap from our calculations is 0.28 eV, in good agreement with previous studies \cite{larson, mishra}. Group theoretical analysis \cite{richter, kohler} of centrosymmetric rhombohedral structure of Bi$_2$Se$_3$ predicts 12 optical zone center phonons to have the symmetry of irreducible representation 2A$_{1g}$+2E$_{g}$+2A$_{2u}$+2E$_{u}$. The atomic vibrations corresponding to E$_{g}$ (A$_{1g}$) symmetry are along (perpendicular) to the layers. Outside the DAC, at ambient pressure four Raman active modes are observed: E$^1_{g}$ at $\sim$ 40 cm$^{-1}$, A$^1_{1g}$ at $\sim$ 72 cm$^{-1}$, E$^2_{g}$ at $\sim$ 130 cm$^{-1}$ and A$^2_{1g}$ at $\sim$ 172 cm$^{-1}$, in agreement with other reports on bulk \cite{richter,gnetz} as well as few quintuples \cite{zha} of Bi$_2$Se$_3$. We could not follow the lowest frequency modes E$^1_{g}$ and A$^1_{1g}$ with pressure inside the DAC due to low signal to noise ratio and high scattering background, and hence will focus our attention on the E$^2_{g}$ and A$^2_{1g}$ modes marked as M1 and M2, respectively, in Fig.~\ref{Fig_1}.

\begin{figure}[p!]
\centering
\includegraphics[trim=0 80 15 40, scale=0.7]{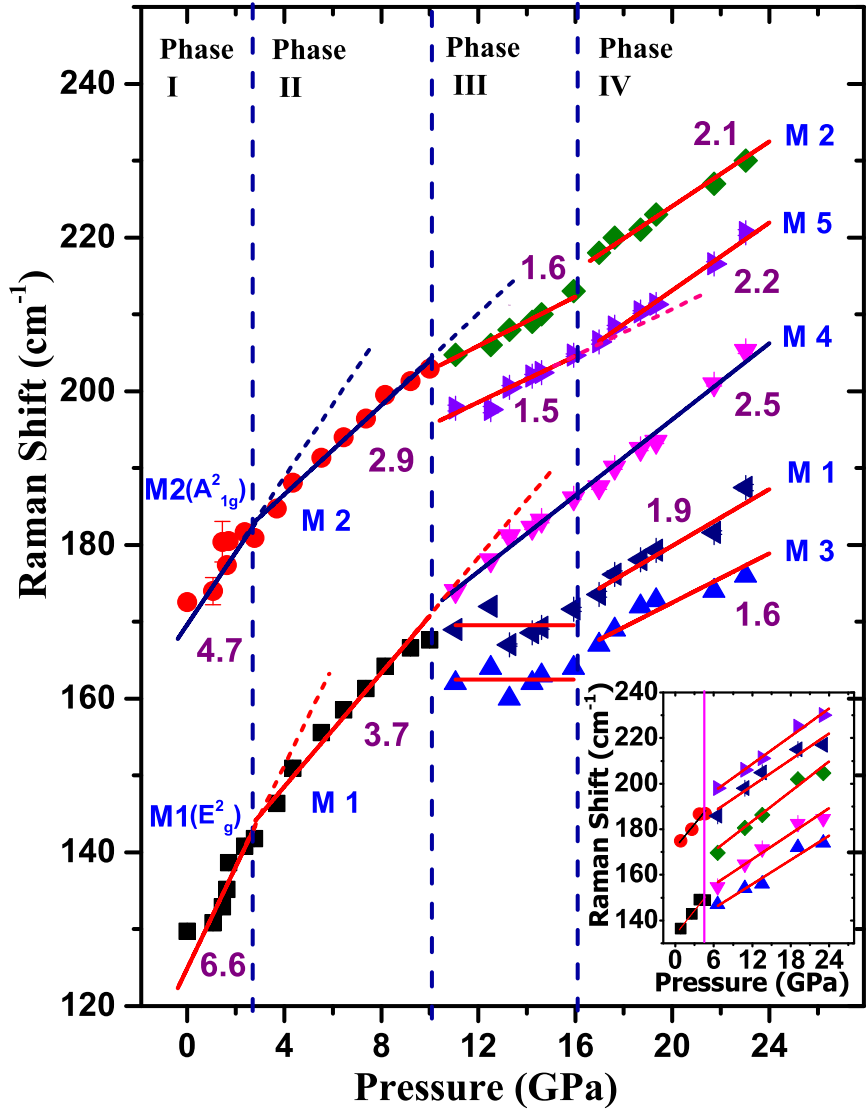}
\caption{(color online)-- Frequency as a function of pressure of various Raman modes. The solid lines are linear fits to the observed frequencies (solid symbols). Error bars (obtained from the fitting procedure) are also shown. The inset represents the observed frequencies in the return pressure run. The vertical (dashed and solid) lines indicate the phase transition pressures. }
\label{Fig_2}
\end{figure}

Fig.~\ref{Fig_1} shows Raman spectra of Bi$_2$Se$_3$ at a few representative elevated pressures. It is clear that beyond 10 GPa, three new modes marked M3 to M5 appear \cite{bs} and continue upto 24 GPa . Beyond 25 GPa, we observe a broad weak peak which almost disappear around 26.2 GPa. The pressure dependence of various Raman modes is shown in Fig.~\ref{Fig_2}. Linear fitting of the experimental data has been done using $\omega_p=\omega_0+(\frac{d\omega}{dP})P$, and the extracted values of the slopes ($\frac{d\omega}{dP}$) are given in Fig.~\ref{Fig_2} as well as in Table I.

From Fig.~\ref{Fig_2}, the following observations can be made: (I) the decrease in pressure coefficient of both M1(E$^2_{g}$) and M2(A$^2_{1g}$) modes occurs at 2.4 GPa; (II) from 10 GPa two phonons M1 and M3 show very little pressure dependence; (III) at 16 GPa, all the four modes except M4 undergo a slope change. We can associate the first transition at P$\sim$ 2.4 GPa as an IST which is consistent with HP X-ray study on Bi$_2$Se$_3$ \cite{bs}, where no discontinuity in unit cell volume was observed. Recently, isostructural transition has been observed \cite{ief} on a similar semiconductor Bi$_2$S$_3$ (of A$_2$B$_3$-type) in the pressure range of 4-6 GPa. Our observations about the change in the slope of high frequency mode A$^2_{1g}$ (M2) at IST is quite contrary to all earlier results of high pressure Raman investigations on Bi$_2$Se$_3$ \cite{bs}, Bi$_2$Te$_3$ \cite{bt} and Sb$_2$Te$_3$ \cite{st}. The change in slope of E$^2_{g}$ mode at IST is larger than that of A$^2_{1g}$.

\begin{table}[htb]
\centering
\caption{Raman modes and pressure coefficients (d$\omega$/dP) observed for various phases.}
\label{tb1}
\begin{tabular}{lccr}
\hline
\hline
\noalign{\smallskip}
%\noalign{\smallskip} \hline  \noalign{\smallskip}
%\noalign{\smallskip} \hline  \noalign{\smallskip}
Phases                    &Mode                    &  Mode  frequency           & d$\omega$/dP \\
                          &                        & $\omega$ (cm$^{-1}$)       & (cm$^{-1}$/GPa)\\

\hline
\noalign{\smallskip}
Phase I(R$\overline{3}$m) &M1(E$^2_{g}$)           &130                &6.6\\
0 to 2.4GPa               &M2(A$^2_{1g}$)          &172                &4.7\\
\noalign{\smallskip}
\hline
\noalign{\smallskip}
Phase II(R$\overline{3}$m) &M1                      &142                &3.7\\
2.4 to 10GPa               &M2                      &181                &2.9\\
\noalign{\smallskip}
\hline
\noalign{\smallskip}
Phase III (C2/m)          &M1                      &169                &$\sim$constant\\
10 to 16GPa               &M2                      &205                &1.6\\
                          &M3(new)                 &162                &$\sim$constant \\
                          &M4(new)                 &173                &2.5\\
                          &M5(new)                 &198                &1.5 \\
\noalign{\smallskip}
\hline
\noalign{\smallskip}
Phase IV (C2/c)           &M1                      &171                &1.9\\
16 to 24GPa               &M2                      &213                &2.1\\
                          &M3                      &164                &1.6 \\
                          &M5                      &205                &2.2 \\
\noalign{\smallskip}
\hline
\hline
%\noalign{\smallskip} \hline  \noalign{\smallskip}
%\noalign{\smallskip} \hline  \noalign{\smallskip}
\end{tabular}
\end{table}

We associate the third observation at $\sim$ 10 GPa with a structural phase transition from rhombohedral ($\alpha$-Bi$_2$Se$_3$) to monoclinic phase ($\beta$-Bi$_2$Se$_3$) in accordance with the HP-XRD results on Bi$_2$Se$_3$ \cite{hamlin, bs}, and also with the theory and experimental work on Bi$_2$Te$_3$ \cite{zhu} and Bi$_2$S$_3$ \cite{ief}. It is clear (from Fig.~\ref{Fig_2}) that like Bi$_2$Te$_3$ \cite{gopal}, we do observe new additional modes around 10 GPa. Group theory predicts a total of 15 Raman active vibrational modes for $\beta$-Bi$_2$Se$_3$ phase, characterized by 10A$_{g}$+5B$_{g}$ \cite{bs,richter, kohler} symmetry. Because of the symmetry lowering, we observe new modes, although we could not observe all the additional Raman modes of lower frequencies. We associate the transition around 16 GPa with a phase transition from $\beta$-Bi$_2$Se$_3$ (C2/m) to $\gamma$-Bi$_2$Se$_3$ (C2/c) structure in accordance with the work on Bi$_2$Te$_3$ \cite{zhu} and also with the recent work on HP Raman of Bi$_2$Se$_3$ \cite{bs}. Above P$\sim$ 24 GPa we observe only a very broad band at $\sim$ 25.3 GPa which becomes even broader at 26.2 GPa in agreement with the earlier results on Bi$_2$Se$_3$ \cite{bs}. We speculate that at P$\sim$ 25 GPa, another phase transition may be occurring from $\gamma$-Bi$_2$Se$_3$ to a disordered bcc structure ($\delta$-Bi$_2$Se$_3$) in analogy with the XRD inference on Bi$_2$Te$_3$ \cite{mari} or to a body-centered tetragonal structure (I4/mmm) as recently shown \cite{jpcm-zhao,sci-yu} . Fig.~\ref{Fig_2} (inset) shows the pressure dependence of all Raman modes in the return pressure cycle, and it is clear that  $\alpha$-Bi$_2$Se$_3$ is recovered around 5 GPa.

\begin{figure}[p!]
\centering
\includegraphics[trim=0 0 40 20, scale=0.8]{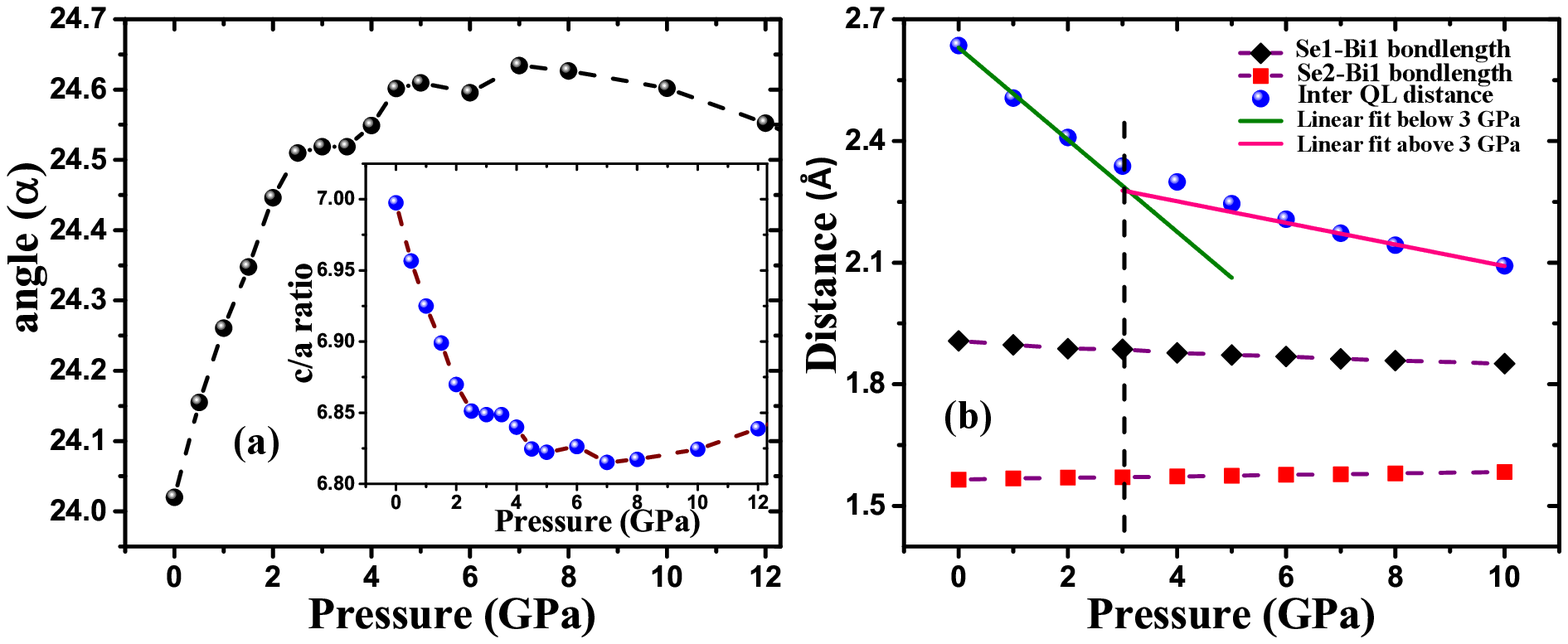}
\caption{(Color online)-- An anomaly in the evolution of structural parameters of Bi$_2$Se$_3$ near P $\sim$ 3 GPa as determined from first-principles calculations. (a) Angle ($\alpha$) of the rhombohedral unit cell increases sharply upto 3 GPa and slowly above it, which is also reflected in the c/a ratio (inset figure). The unusual behaviour of the $\alpha$ or c/a ratio appears due to the sharp decease of the inter quintuple layer (QL) distances below 3 GPa as evident in figure (b). Change in slope in the graph is marked with vertical dashed lines. }
\label{Fig_3}
\end{figure}

We have carried out total energy calculations of the three different phases ($\alpha$, $\beta$ and $\gamma$) of Bi$_2$Se$_3$ as a function of pressure. Enthalpies H(P) of these structures reveal that $\alpha$ to $\beta$ phase transition occurs at 12 GPa in agreement with our experiment as well as others \cite{bs,jpc-liu,jpcm-zhao} and $\beta$ to $\gamma$ phase transition takes place at 28 GPa (see Figs. S1 and S2 in supplementary material). The difference in our experimental transition pressure from $\beta$ to $\gamma$ phase at 16 GPa and calculated transition pressure of $\sim$ 28 GPa may be related to some non-hydrostatic component in the experiments, as also mentioned by Vilaplana et al. \cite{bs} in explaining their results.

We now analyze the nature of the phase transition at 2.4 GPa in details through the results of first-principles calculations. Our calculation show an unusual change in the internal angle ($\alpha$) of the rhombohedral unit cell near 3 GPa (see Fig.~\ref{Fig_3}(a)). This anomalous change is also reflected in an anomaly in the c/a ratio of Bi$_2$Se$_3$ (see inset of Fig.~\ref{Fig_3}a) near that pressure. To find the origin of this change in $\alpha$ (or c/a ratio) we examine the intra and inter QL distances between atomic planes (see Fig.~\ref{Fig_3}b), and we find that it is the inter-quintuple layer (QL) distance (a$_{QL}$) which is responsible for the anomaly in the c/a ratio near 3 GPa. There is a distinct change in the da$_{QL}$/dP around 3 GPa (see Fig.~\ref{Fig_3}b), which can be responsible for the observed changes in the slope of E$^2_{g}$ and A$^2_{1g}$ Raman frequencies at this pressure. \\

\begin{figure}[p!]
\centering
\includegraphics[trim=0 0 40 0, scale=0.8]{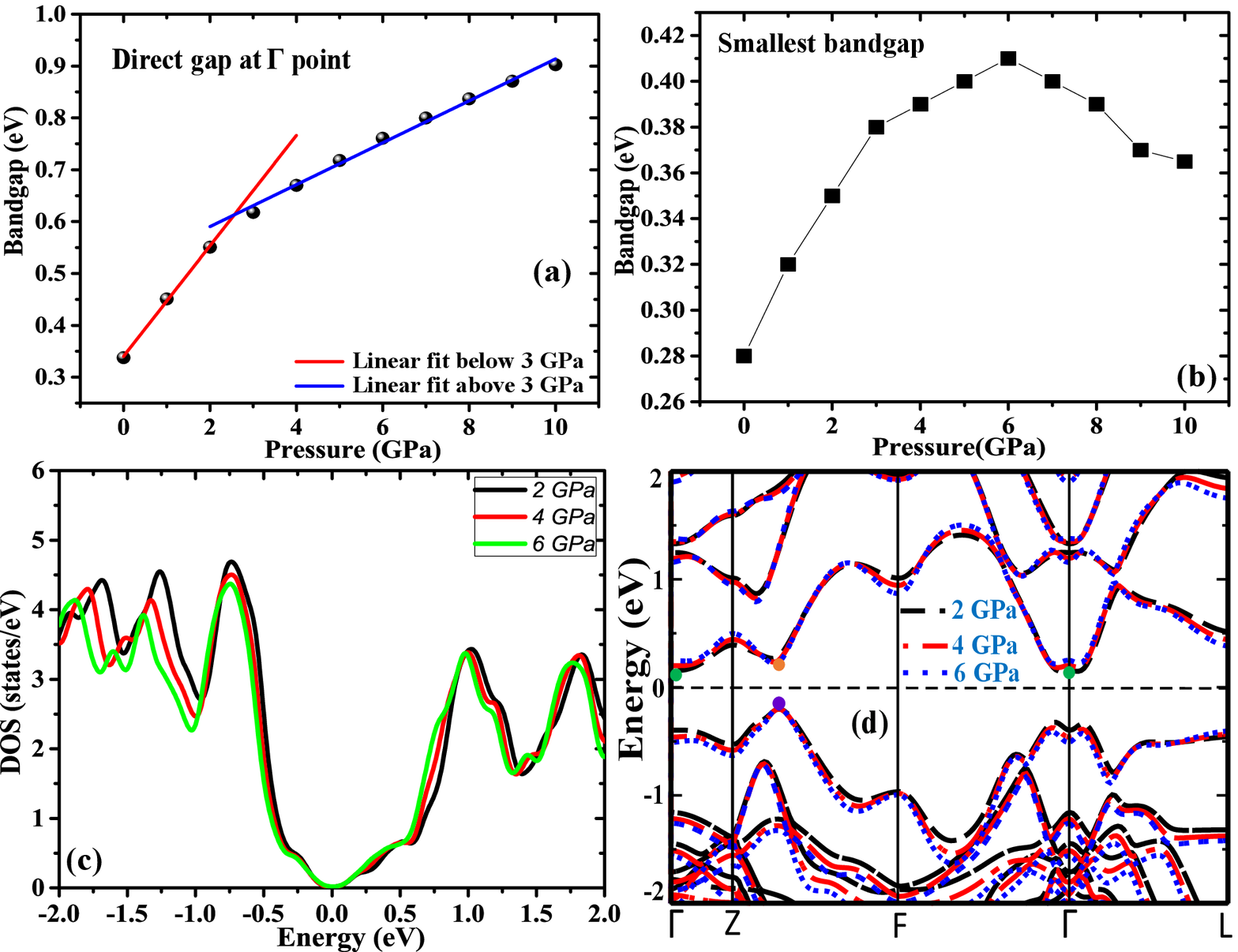}
\caption{(Color online)--Calculated electronic structure as a function of pressure. (a) Change in the slope of direct band gap of Bi$_2$Se$_3$ at the $\Gamma$ point near P $\sim$ 3 GPa, which is a consequence of the unusual change of c/a ratio near that pressure. (b) Smallest band gap as a function of pressure and (c) total electronic density of states at 2 GPa, 4 GPa and 6 GPa. (d) Though the band gap at $\Gamma$-point of Bi$_2$Se$_3$ increases with pressure, its electronic structure reveals small changes in the extrema in the bands around 4 GPa. Closer inspection of electronic structures reveals an indirect band gap of Bi$_2$Se$_3$ at 2 GPa, where VBM and CBM are marked with violet and green dots respectively, whereas at 6 GPa, Bi$_2$Se$_3$ exhibits a direct band gap where VBM and CBM are denoted with violet and orange dots respectively. We note that, with pressure the VBM remains fixed, but CBM changes its positions from $\Gamma$-point (green dot) to along Z-F line (orange dot).}
\label{Fig_4}
\end{figure}

It is clear from all the earlier experiments that anomalies in phonon spectrum are present around $\sim$ 3 GPa. To relate its relevance to ETT, we have performed detailed calculations on the bulk band gap as well as surface electronic structure below and above the transition region ($\sim$ 3 GPa). The direct band gap of the bulk of Bi$_2$Se$_3$ at $\Gamma$ point increases with pressure and a change in the slope of the band gap appears near 3 GPa (see Fig.~\ref{Fig_4}a). Though the direct band gap at $\Gamma$ point increases monotonically with pressure, the smallest band gap of the bulk first increases upto 6 GPa and decreases above it (see Fig.~\ref{Fig_4}b). With increasing pressure, the valence band maxima (VBM) and conduction band minima (CBM) change their positions in the momentum space making Bi$_2$Se$_3$ an indirect \cite{hzhang} to direct band gap material above P $>$ 4 GPa \cite{PRBR-Nech}. Electronic density of states (see Fig.~\ref{Fig_4}c) reveal a small change in the band gap of Bi$_2$Se$_3$ as shown for a few typical pressures of 2 GPa, 4 GPa and 6 GPa. The electronic structure of the bulk calculated by including spin-orbit coupling shows a gap of 0.28 eV at ambient conditions. We compare the electronic structure of the bulk at different pressures near this transition (see Fig.~\ref{Fig_4}d) and we do not find any significant changes in the electronic band dispersions. As  Bi$_2$Se$_3$ is a strong $\mathbb{Z}_2$  topological insulator with topological invariant $\nu_0$=1, we determined the $\mathbb{Z}_2$ invariant $\nu_0$ of Bi$_2$Se$_3$ at pressures upto 8 GPa following the method developed by Fu and Kane \cite{fu} for a centrosymmetric material. In this method, we find the parity of the occupied bands at eight time reversal invariant momenta (TRIM), and use the relation (-1)$^{\nu_0}$ = $\Pi_{i=1}^{8}$ $\delta_i$, where i runs over eight TRIM and $\delta_i$=$\Pi_{m}$ $\xi^{i}_{2m}$, $\xi_{2m}$  being the parity of the occupied bands indexed with 2m at each TRIM (i). We find that $\nu_0$ remains 1 at all pressures upto 8 GPa (see Table II). This signifies that there is no electronic topological transition occurring in the above pressure range and Bi$_2$Se$_3$ remains a topological insulator before it undergoes a structural transition to monoclinic structure at higher pressure.\\

\begin{figure}[p!]
\centering
\includegraphics[trim=0 0 40 20, scale=0.8]{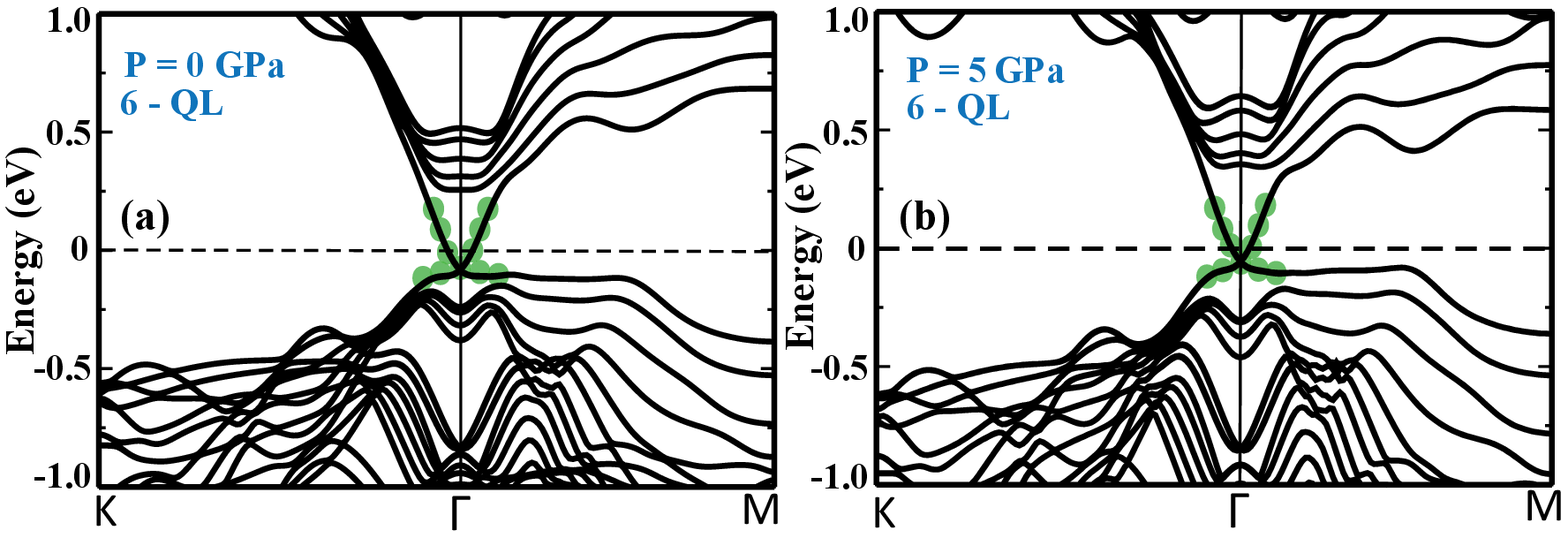}
\caption{(Color online)--Electronic structure of the surface of Bi$_2$Se$_3$ for (a) 0 GPa and (b) 5 GPa which reveal of a single Dirac cone (highlighted with green dots) at the $\Gamma$ point  showing
no change in robust bulk electronic topology (which gives rise to gapless surface Dirac cone) as a function of pressure.}
\label{Fig_5}
\end{figure}

As gapless Dirac cone in the surface electronic structure is characteristic of a strong topological insulator, we calculate the electronic structure of the (001) surface of Bi$_2$Se$_3$ as a function of pressure to see any topological change, as it is the non-trivial topology of the bulk electronic wave function which gives rise to symmetry protected Dirac cone on the surface. At ambient pressure, we find a gapless Dirac cone (Fig.~\ref{Fig_5}a) as expected of a strong $\mathbb{Z}_2$ TI. Our first-principles calculation of the slab of Bi$_2$Se$_3$ reveals that, on the other side of the transition (e.g. at P=5 GPa), band gap does not open up at $\Gamma$ in its surface electronic spectrum maintaining its topological insulating nature intact (see Fig.~\ref{Fig_5}b), in agreement with the observed angle-resolved photoemission spectroscopy (ARPES) \cite{yzhang2}. Thus, there is clearly no change in the $\mathbb{Z}_2$ topological invariant of the electronic structure of Bi$_2$Se$_3$ through the transition at P$\sim$ 3 GPa.

\begin{figure}[p!]
\centering
\includegraphics[trim=0 0 40 20, scale=0.8]{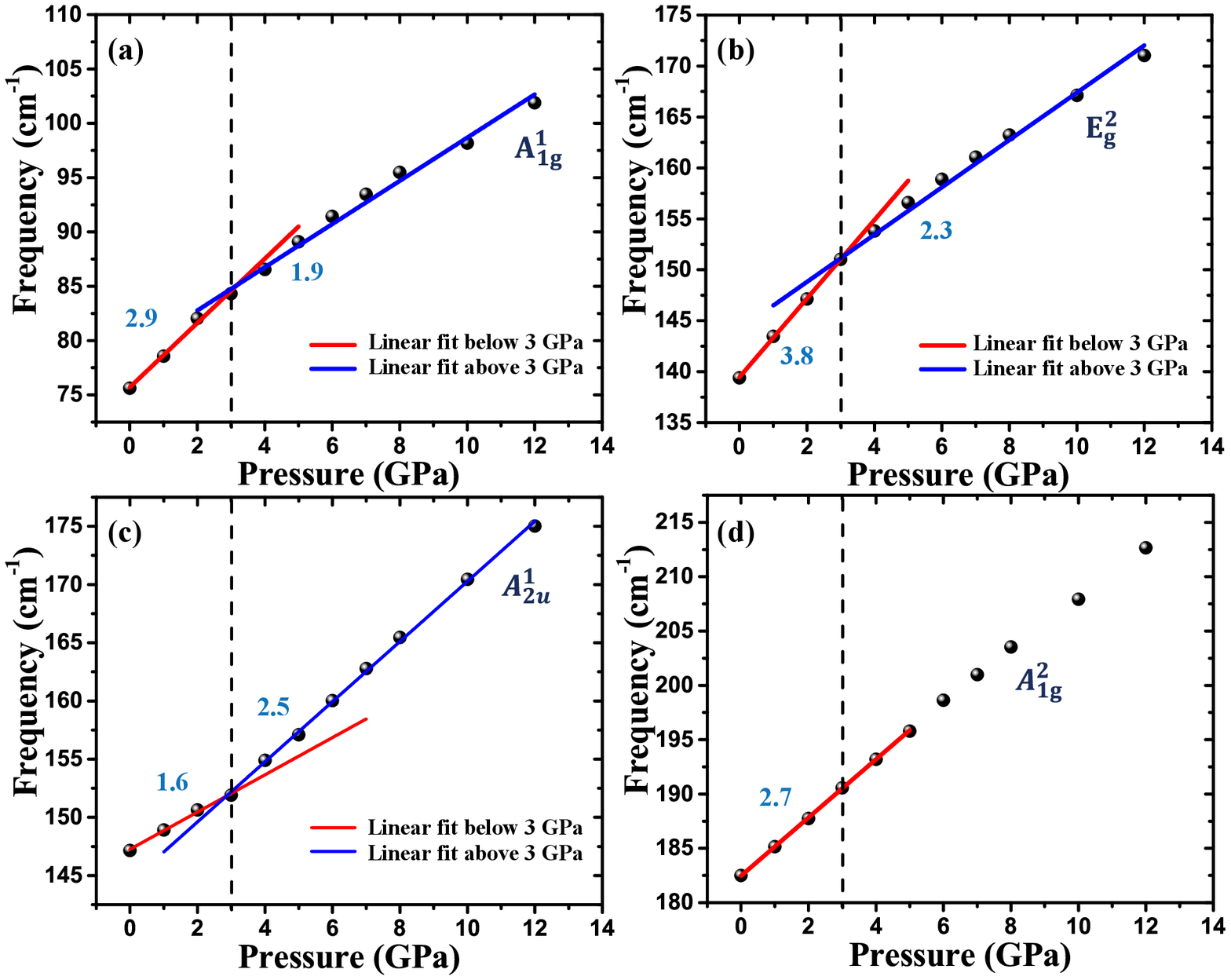}
\caption{(Color online)--Calculated Raman and IR active modes as a function of pressure. The change of slope of the phonon modes as measured by the Raman experiment is also captured through our first-principles calculation. Change in slope (expressed in cm$^{-1}$/GPa) for Raman active modes A$_{1g}^1$, A$_{1g}^2$, E$_g^2$ are shown in (a), (d) and (b) respectively. Infrared active mode A$_{2u}^1$ also displays a change in slope  near 3 GPa shown in (c). Vertical dashed lines are placed at the positions where the change in slope is observed. }
\label{Fig_6}
\end{figure}

In the light of recent results of Forster et al \cite{prb-forster}, we note that we have determined the topological invariants of Bi$_2$Se$_3$ from calculations of its bulk form. Errors in the band gap in a density functional theoretical (DFT) calculations may result in some errors in the pressure of transition from band to topological insulating states. On the other hand, closure of a gap in the surface electronic structure is sensitive to the error in the DFT band gap, but it is a finite size effect. Topological character can always be ascertained by using a thicker film or slab in the calculations.

Now, we  discuss the calculated vibrational frequencies of Bi$_2$Se$_3$ as a function of pressure which, do reveal a change in their slopes (see Fig.~\ref{Fig_6}) near the transition (i.e. in 2-5 GPa range of pressure) where the rhombohedral angle ($\alpha$), c/a ratio and inter QL distance exhibit anomalous behaviour (see Fig.3). The most significant changes occur in the Raman active  $E_g^{2}$ and $A_{1g}^1$ modes and also in the IR active $A_{2u}^1$ mode. The change in slope of the A$^2_{1g}$ mode is not captured in the calculation due to the anharmonic coupling between the A$^2_{1g}$ and all other modes and hence pressure dependence of this mode originating from strong anharmonicity is not captured within the harmonic analysis presented here.\\

\begin{table}[htb]
\centering
\label{t:parity}
\scalebox{0.8}{
\begin{tabular}{c|c||cccccccccccccccc||c}%{l@{\hskip 0.1in}c@{\hskip 0.1in}c@{\hskip 0.1in}c@{\hskip 0.1in}c@{\hskip 0.1in}c@{\hskip 0.1in}c@{\hskip 0.1in}c@{\hskip 0.1in}c}

\noalign{\smallskip} \hline  \noalign{\smallskip}

 &  P (0 GPa)   & + & - & + & - & + & - & + & + & - & + & - & - & - & + & ; & - & (-1) \\
Bi$_2$Se$_3$ & P(4 GPa) & + & - & + & - & + & - & + & + & - & + & - & - & - & + & ; & - & (-1) \\
 &  P (8 GPa)   & + & - & + & - & + & - & + & + & - & + & - & - & - & + & ; & - & (-1) \\

\noalign{\smallskip} \hline  \noalign{\smallskip}
\end{tabular}   }
\caption {Parities of the fourteen occupied bands and the lowest energy unoccupied band of Bi$_2$Se$_3$ at pressures in the range of 0-8 GPa. The product of parities of the valence band manifold are -1 as indicated within the brackets, which gives the value of $\mathbb{Z}_2$-invariant $\nu_0$ to be 1.}
\end{table}

\section{Conclusions}
\label{sec:Conclusion}

In summary, our high pressure Raman experiments reveal changes in pressure coefficients (d$\omega$/dP) of the A$^2_{1g}$ and $E_g^{2}$ modes around the isostructural transition at P$\sim$3 GPa. Our theoretical calculations confirm the change in slope of $E_g^{2}$ mode, while the change in slope of A$^2_{1g}$ mode at the isostructural transition needs further understanding. Our calculations clearly show that the low pressure transition at $\sim$ 3 GPa is $\textit{not}$ related to any change in the electronic topology, as there is no change in the $\mathbb{Z}_2$ index, and the Dirac cone in the surface electronic structure remains intact below and above the transition. Hence, the lowest pressure transition should be better termed as an isostructural transition, and not an ETT.

\section{Acknowledgments}

AKS acknowledges the funding from Department of Science and Technology, India. AKS and UVW acknowledge funding from  JC Bose National Fellowships. AB thanks CSIR for a research fellowship. \\


\begin{thebibliography}{70}

\bibitem{yando}
Y. Ando, Journal of the Physical Society of Japan {\bf 82},  102001  (2013).

\bibitem{mfru}
M. Fruchart and D. Carpentier, Comptes Rendus Physique {\bf 14},  779  (2013).

\bibitem{moore}
J.~E. Moore, Nature (London) {\bf 464},  194  (2010).

\bibitem{hasan}
M.~Z. Hasan and C.~L. Kane, Rev. Mod. Phys. {\bf 82},  3045  (2010).

\bibitem{qi}
X.-L. Qi and S.-C. Zhang, Physics Today {\bf 63},  33  (2010).

\bibitem{fu}
L. Fu and C.~L. Kane, Phys. Rev. B {\bf 76},  045302  (2007).

\bibitem{mele}
C.~L. Kane and E.~J. Mele, Phys. Rev. Lett. {\bf 95},  146802  (2005).

\bibitem{hug}
B.~A. Bernevig, T.~L. Hughes, and S.-C. Zhang, Science {\bf 314},  1757
  (2006).

\bibitem{dzi}
P. Dziawa {\it et~al.}, Nat Mater {\bf 11},  1023  (2012).

\bibitem{yang}
Y. Yang {\it et~al.}, Applied Physics Letters {\bf 99},  182101  (2011).

\bibitem{wang}
Y.~H. Wang {\it et~al.}, Phys. Rev. Lett. {\bf 107},  207602  (2011).

\bibitem{zutic}
I. \ifmmode \check{Z}\else \v{Z}\fi{}uti\ifmmode~\acute{c}\else \'{c}\fi{}, J.
  Fabian, and S. Das~Sarma, Rev. Mod. Phys. {\bf 76},  323  (2004).

\bibitem{nayak}
C. Nayak {\it et~al.}, Rev. Mod. Phys. {\bf 80},  1083  (2008).

\bibitem{more}
J. Moore, Nat Phys {\bf 5},  378  (2009).

\bibitem{hong}
S.~S. Hong {\it et~al.}, Nano Letters {\bf 10},  3118  (2010).

\bibitem{zha}
J. Zhang {\it et~al.}, Nano Letters {\bf 11},  2407  (2011).

\bibitem{yzhang2}
Y. Zhang {\it et~al.}, Nat Phys {\bf 6},  584  (2010).

\bibitem{hstein1}
H. Steinberg, D.~R. Gardner, Y.~S. Lee, and P. Jarillo-Herrero, Nano Letters
  {\bf 10},  5032  (2010).

\bibitem{kim1}
D. Kim {\it et~al.}, Nat Phys {\bf 8},  459  (2012).

\bibitem{kim2}
D. Kim {\it et~al.}, Nano Letters {\bf 14},  1701  (2014).

\bibitem{hzhang}
H. Zhang {\it et~al.}, Nat Phys {\bf 5},  438  (2009).

\bibitem{pal2014strain}
K. Pal and U.~V. Waghmare, Applied Physics Letters {\bf 105},  062105  (2014).

\bibitem{PRL1}
A. Bera {\it et~al.}, Phys. Rev. Lett. {\bf 110},  107401  (2013).

\bibitem{steve}
S.~M. Young {\it et~al.}, Phys. Rev. B {\bf 84},  085106  (2011).

\bibitem{liu1}
W. Liu {\it et~al.}, Phys. Rev. B {\bf 84},  245105  (2011).

\bibitem{einaga}
M. Einaga {\it et~al.}, Journal of Physics: Conference Series {\bf 215},
  012036  (2010).

\bibitem{jl}
J.~L. Zhang {\it et~al.}, Proceedings of the National Academy of Sciences {\bf
  108},  24  (2011).

\bibitem{czhang1}
C. Zhang {\it et~al.}, Phys. Rev. B {\bf 83},  140504  (2011).

\bibitem{andy}
A. Das {\it et~al.}, Nat Phys {\bf 8},  887  (2012).

\bibitem{ewang}
E. Wang {\it et~al.}, Nat Phys {\bf 9},  621  (2013).

\bibitem{svovs1}
S.~V. Ovsyannikov {\it et~al.}, Journal of Applied Physics {\bf 104},  053713
  (2008).

\bibitem{hamlin}
J.~J. Hamlin {\it et~al.}, Journal of Physics: Condensed Matter {\bf 24},
  035602  (2012).

\bibitem{bs}
R. Vilaplana {\it et~al.}, Phys. Rev. B {\bf 84},  184110  (2011).

\bibitem{bt}
R. Vilaplana {\it et~al.}, Phys. Rev. B {\bf 84},  104112  (2011).

\bibitem{st}
O. Gomis {\it et~al.}, Phys. Rev. B {\bf 84},  174305  (2011).

\bibitem{poli}
A. Polian {\it et~al.}, Phys. Rev. B {\bf 83},  113106  (2011).

\bibitem{lif}
M. Lifshitz, Sov. Phys. JETP {\bf 11},  1130  (1960).

\bibitem{gopal}
G.~K. Pradhan {\it et~al.}, Solid State Communications {\bf 152},  284  (2012).

\bibitem{PSSB_Review}
F.~J. Manjón {\it et~al.}, physica status solidi (b) {\bf 250},  669  (2013).

\bibitem{PRL_support}
K. Saha, K. L\'egar\'e, and I. Garate, Phys. Rev. Lett. {\bf 115},  176405
  (2015).

\bibitem{GIANOZZI}
Quantum-ESPRESSO is a community project for high-quality quantum-simulation
  soft- ware, based on density-functional theory, and coordinated by P.
  Giannozzi. See http://www. quantum-espresso.org and http://www.pwscf.org  .

\bibitem{baroni}
S. Baroni, S. de~Gironcoli, A. Dal~Corso, and P. Giannozzi, Rev. Mod. Phys.
  {\bf 73},  515  (2001).

\bibitem{HUA}
X. Hua, X. Chen, and W.~A. Goddard, Phys. Rev. B {\bf 55},  16103  (1997).

\bibitem{PERDEW}
J.~P. Perdew, K. Burke, and M. Ernzerhof, Phys. Rev. Lett. {\bf 77},  3865
  (1996).

\bibitem{PACK}
H.~J. Monkhorst and J.~D. Pack, Phys. Rev. B {\bf 13},  5188  (1976).

\bibitem{wise}
J. Wiese and L. Muldawer, Journal of Physics and Chemistry of Solids {\bf 15},
  13  (1960).

\bibitem{nakajima}
S. Nakajima, Journal of Physics and Chemistry of Solids {\bf 24},  479
  (1963).

\bibitem{black}
J. Black, E. Conwell, L. Seigle, and C. Spencer, Journal of Physics and
  Chemistry of Solids {\bf 2},  240   (1957).

\bibitem{larson}
P. Larson {\it et~al.}, Phys. Rev. B {\bf 65},  085108  (2002).

\bibitem{mishra}
S.~K. Mishra, S. Satpathy, and O. Jepsen, Journal of Physics: Condensed Matter
  {\bf 9},  461  (1997).

\bibitem{richter}
W. Richter and C.~R. Becker, physica status solidi (b) {\bf 84},  619  (1977).

\bibitem{kohler}
H. Köhler and C.~R. Becker, physica status solidi (b) {\bf 61},  533  (1974).

\bibitem{gnetz}
V. Gnezdilov {\it et~al.}, Phys. Rev. B {\bf 84},  195118  (2011).

\bibitem{ief}
I. Efthimiopoulos {\it et~al.}, The Journal of Physical Chemistry A {\bf 118},
  1713  (2014).

\bibitem{zhu}
L. Zhu {\it et~al.}, Phys. Rev. Lett. {\bf 106},  145501  (2011).

\bibitem{mari}
M. Einaga {\it et~al.}, Phys. Rev. B {\bf 83},  092102  (2011).

\bibitem{jpcm-zhao}
J. Zhao {\it et~al.}, Journal of Physics: Condensed Matter {\bf 25},  125602
  (2013).

\bibitem{sci-yu}
Z. Yu {\it et~al.}, Scientific Reports {\bf 5},  15939  (2015).

\bibitem{jpc-liu}
G. Liu {\it et~al.}, The Journal of Physical Chemistry C {\bf 117},  10045
  (2013).

\bibitem{PRBR-Nech}
I.~A. Nechaev {\it et~al.}, Phys. Rev. B {\bf 87},  121111  (2013).

\bibitem{prb-forster}
T. F\"orster, P. Kr\"uger, and M. Rohlfing, Phys. Rev. B {\bf 92},  201404
  (2015).

\end{thebibliography}
\end{document}